\documentstyle[12pt,fleqn]{article}
\input epsf
\def\fun#1#2{\lower3.6pt\vbox{\baselineskip0pt\lineskip.9pt
        \ialign{$\mathsurround=0pt#1\hfill##\hfil$\crcr#2\crcr\sim\crcr}}}
\begin{document}

\setcounter{page}{0}
\thispagestyle{empty}

\begin{flushright}
{\footnotesize
FERMILAB--PUB--98/154--A\\
CERN-TH/98-160\\
OUTP-98-40P\\
hep-ph/9805473\\
May 1998\\}
\end{flushright}

\vskip .5cm

\begin{center}
{\Large\bf  Nonthermal Supermassive Dark Matter }

\vskip 1cm

{\bf 
Daniel J. H. Chung,$^{a,b,}$\footnote{E-mail: 
                         {\tt djchung@theory.uchicago.edu}}
Edward W. Kolb,$^{b,c,}$\footnote{E-mail: 
                         {\tt  rocky@rigoletto.fnal.gov}}
Antonio Riotto$^{d,}$\footnote{E-mail: 
                         {\tt riotto@nxth04.cern.ch}}$^,$\footnote{On 
                          leave  from Department of Theoretical Physics,
                          University of Oxford, U.K. }}
\vskip .75cm
{\it 
$^a$Department of Physics and Enrico Fermi Institute \\ 
The University of Chicago, Chicago, Illinois 60637-1433\\
\vspace{12pt}
$^b$NASA/Fermilab Astrophysics Center \\ Fermilab
National Accelerator Laboratory, Batavia, Illinois~~60510-0500\\
\vspace{12pt}
$^c$Department of Astronomy and Astrophysics and  Enrico Fermi Institute\\
The University of Chicago, Chicago, Illinois~~60637-1433\\
\vspace{12pt}
$^d$Theory Division, CERN, CH-1211 Geneva 23, Switzerland
}
\end{center}
\vskip .5cm
\baselineskip=24pt

\begin{quote}
\hspace*{2em} We discuss several cosmological production mechanisms
for nonthermal supermassive dark matter and argue that dark matter may
be elementary particles of mass much greater than the weak scale.
Searches for dark matter should not be limited to weakly interacting
particles with mass of the order of the weak scale, but should extend
into the supermassive range as well.
\vspace*{8pt}

PACS number(s): 98.80.Cq

\end{quote}

\renewcommand{\thefootnote}{\arabic{footnote}}
\addtocounter{footnote}{-3}

\newpage

\setcounter{page}{1}

\def\simlt{\stackrel{<}{{}_\sim}}
\def\simgt{\stackrel{>}{{}_\sim}}

There is conclusive evidence that the dominant component of the matter
density in the universe is dark. The most striking indication of the
existence of dark matter (DM) is the observations of flat rotation
curves for spiral galaxies \cite{vel}, indicating that DM in galactic
halos is about ten times more abundant than the luminous component.
Dynamical evidence for DM in clusters of galaxies is also compelling.
In terms of the critical density, $\rho_C=3 H_0^2 M_{{\rm Pl}}^2/8\pi$
with $H_0\equiv 100\: h \:{\rm km}\:{\rm sec}^{-1}\: {\rm Mpc}^{-1}$
and $M_{{\rm Pl}}$ the Planck mass, the amount of DM inferred from
dynamics of clusters of galaxies is $\Omega_{{\rm DM}}\equiv
\rho_{{\rm DM}}/\rho_C\simgt 0.3$. In addition, the most natural
inflation models predict a flat universe, {\it i.e.,} $\Omega_0=1$,
while standard big-bang nucleosynthesis implies that ordinary baryonic
matter can contribute at most $10\%$ to $\Omega_0$.  This means that
about $ 90 \%$ of the matter in our universe may be dark.

It is usually assumed that DM consists of a species of a new, yet
undiscovered, massive particle we denote as $X$.  It is also often
assumed that the DM is a thermal relic, {\it i.e.,} it was in chemical
equilibrium in the early universe.  

The simple assumption that the DM is a thermal relic is surprisingly
restrictive.  The limit $\Omega_X \simlt 1$ implies that the mass of a
DM relic must be less than about 500 TeV \cite{griestkam}. This upper
bound turns out to be fatal to the proposal that DM consists of
charged massive particles (CHAMPs---${\cal C}^{\pm}$) \cite{champ}.
The present experimental limits on superheavy hydrogen and ${\cal
C}^{-}p$ atoms are compatible with the CHAMP scenario only if they are
more massive than about $10^{3}$ TeV \cite{limit}.  Similarly, current
limits from underground detectors exclude the possibility that halo DM
consists of {\it colored} particles of mass less than 500 TeV. The
standard lore is that the hunt for DM should concentrate on particles
with mass of the order of the weak scale and with interaction with
ordinary matter on the scale of the weak force. This has been the
driving force behind the vast effort in DM detectors.

In view of the unitarity argument, in order to consider {\it thermal}
supermassive dark matter,\footnote{In this paper, {\it supermassive}
implies much more massive than the weak scale (about 100 GeV).} one
must invoke, for example, late-time entropy production to dilute the
abundance of these supermassive particles \cite{k}, rendering the
scenario unattractive.

In this Letter we argue that recent developments in understanding how
matter is created in the early universe suggests the possibility that
that DM in the Universe might be naturally composed of {\it
nonthermal} supermassive states.  The supermassive dark matter (SDM)
particles $X$ may have a mass possibly as large as the Grand Unified
Theory (GUT) scale. We suggest a number of cosmological production
mechanisms for nonthermal supermassive dark matter. It is very
intriguing that our considerations resurrect the possibility that the
dark matter might be charged or even strongly interacting!

We discuss four production mechanisms.  We first propose production
during reheating of the universe after inflation.  We point out that
if the reheat temperature is denoted as $T_{RH}$, the present
abundance of SDM is proportional to $(2000 M_X/T_{RH})^{-7}$, rather
than $\exp(-M_X/T_{RH})$ as one might naively expect.  We then suggest
the possibility of SDM production in preheating, making use of
previous work that considered production of massive particles for
baryogenesis.  We then review the possibility of gravitational
production of SDM at the end of the inflationary era.  Finally, we
propose that SDM might be created in the collisions of vacuum bubbles
in a first-order phase transition.

There are two necessary conditions for an SDM scenario.  First, the
SDM must be stable, or at least have a lifetime greater than the age
of the universe.  This may result from, for instance, supersymmetric
theories where the breaking of supersymmetry is communicated to
ordinary sparticles via the usual gauge forces \cite{review}. In
particular, the secluded and the messenger sectors often have
accidental symmetries analogous to baryon number. This means that the
lightest particle in those sectors might be stable and very massive if
supersymmetry is broken at a large scale \cite{raby}.  Other natural
candidates for supermassive DM arise in theories with discrete gauge
symmetries \cite{discrete} and in string theory and M theory
\cite{john}. In the M-theory case, stable or metastable bound states
of matter in the hidden sector, called cryptons, seem to be favoured
over other possible candidates in string or M theory, such as the
Kaluza-Klein states associated with extra dimensions. A specific
string model that predicts cryptons as hidden-sector bound states
weighing about $10^{12}$ GeV is exhibited in \cite{john}.

The second condition for SDM is that the particle must not have been
in equilibrium when it froze out ({\it i.e.,} it is not a thermal
relic), otherwise $\Omega_X$ would be larger than one.\footnote{In
this paper $\Omega_i$ refers to the {\it present} value for species
$i$} A sufficient condition for nonequilibrium is that the
annihilation rate (per particle) must be smaller than the expansion
rate: $n\sigma|v|<H$, where $n$ is the number density, $\sigma |v|$ is
the annihilation rate times the M{\o}ller flux factor, and $H$ is the
expansion rate.  Conversely, if the SDM was created at some
temperature $T_*$ {\it and} $\Omega_X<1$, then it is easy to show that
it could not have attained equilibrium.  To see this, assume $X$'s
were created in a radiation-dominated universe at temperature $T_*$.
Then $\Omega_X$ is given by $\Omega_X =
\Omega_\gamma(T_*/T_0)m_Xn_X(T_*)/\rho_\gamma(T_*)$, where $T_0$ is
the present temperature.  (In this paper we will ignore dimensionless
factors of order unity.)  Using the fact that $\rho_\gamma(T_*) =
H(T_*) M_{Pl} T_*^2$, we find $n_X(T_*)/H(T_*) =
(\Omega_X/\Omega_\gamma) T_0 M_{Pl} T_*/M_X$.  Since we may safely take the
limit $\sigma |v| < M_X^{-2}$, $n_X(T_*)\sigma |v| / H(T_*)$ must
be less than $(\Omega_X / \Omega_\gamma) T_0 M_{Pl} T_* / M_X^3$.
Thus, the requirement for nonequilibrium is
\begin{equation}
\left( \frac{200\,{\rm TeV}}{M_X}\right)^2 \left( \frac{T_*}{M_X}
\right) < 1 \ .
\end{equation}
This implies that if a nonrelativistic particle with $M_X\simgt 200$
TeV was created at $T_*<M_X$ with a density low enough to result in
$\Omega_X \simlt 1$, then its abundance must have been so small that
it never attained equilibrium. Therefore, if there is some way to
create SDM in the correct abundance to give $\Omega_X\sim 1$,
nonequilibrium is guaranteed.

An attractive origin for SDM is during the defrosting phase after
inflation.  It is important to realize that it is not necessary to
convert a significant fraction of the available energy into massive
particles; in fact, it must be an infinitesimal amount.  If a fraction
$\epsilon$ of the available energy density is in the form of a
massive, stable $X$ particle, then $\Omega_X = \epsilon \Omega_\gamma
(T_{RH}/T_0)$, where $T_{RH}$ is the ``reheat'' temperature.  For
$\Omega_X=1$, this leads to the limit $\epsilon \simlt 10^{-17}
(10^9\, {\rm GeV}/T_{RH})$.  We will discuss how particles of mass
much greater than $T_{RH}$ may be created after inflation.

In one extreme we might assume that the vacuum energy of inflation is
immediately converted to radiation resulting in a reheat temperature
$T_{RH}$.  In this case $\Omega_X $ can be calculated by integrating
the Boltzmann equation with initial condition $N_X=0$ at $T=T_{RH}$.
One expects the $X$ density to be suppressed by $\exp(-2M_X/T_{RH})$;
indeed, one finds $\Omega_X \sim 1$ for $M_X/T_{RH} \sim 25 +
0.5\ln(m_X^2\langle \sigma |v|\rangle)$, in agreement with previous
estimates \cite{vadimvaleri} that for $T_{RH}\sim10^9$GeV, the SDM
mass would be about $2.5\times10^{10}$GeV.

A second (and more plausible) scenario is that reheating is not
instantaneous, but is the result of the decay of the inflaton field.
In this approach the radiation is produced as the inflaton decays.
The SDM density is found by solving the coupled system of equations
for the inflaton field energy, the radiation density, and the SDM mass
density.  The calculation has been recently reported in Ref.\
\cite{CKRII}, with the result $\Omega_X \sim m_X^2\langle \sigma
|v|\rangle (2000 T_{RH}/M_X)^7$.  Note that the suppression in
$M_X/T_{RH}$ is not exponential, but a power law (albeit a large
power).  Another crucial feature is the rather large factor of $10^4$.
This implies that for a reheat temperature as low as $10^9$GeV, a
particle of mass $10^{13}$GeV can be produced in sufficient abundance
to give $\Omega_X \sim 1$.

The large difference in SDM masses in the two reheating scenarios
arises because the peak temperature is much larger in the second
scenario, even with identical $T_{RH}$.  Because the temperature
decreases as $a^{-3/8}$ ($a$ is the scale factor) during most of the
reheating period in the second scenario, it must have once been much
greater than $T_{RH}$.  If we assume the radiation spectrum did not
depart grossly from thermal, the effective temperature having once
been larger than $T_{RH}$ implies that the density of particles with
enough energy to create SDM was larger.  Denoting as $T_2$ the maximum
effective temperature for the second scenario, $T_2/T_{RH} \sim
(M_\phi/\Gamma_\phi)^{1/4} \gg 1$, where $\Gamma_\phi$ is the
effective decay rate of the inflaton.  See \cite{CKRII} for details.

Another way to produce SDM after inflation is in a preliminary stage
of reheating called ``preheating'' \cite{explosive}, where nonperturbative
quantum effects may lead to an extremely effective dissipational
dynamics and explosive particle production. Particles can be created
in a broad parametric resonance with a fraction of the energy stored
in the form of coherent inflaton oscillations at the end of inflation
released after only a dozen oscillation periods.  A crucial
observation for our discussion is that particles with mass up to
$10^{15}$ GeV may be created during preheating \cite{klr,kt,gut}, and
that their distribution is nonthermal. If these particles are stable,
they may be good candidates for SDM.

To study how the creation of SDM occurs in preheating, let's
take the simplest chaotic inflation potential:
$V(\phi)=M_\phi^2\phi^2/2$ with $M_\phi\sim 10^{13}$ GeV.  We assume
that the interaction term between the SDM and the inflaton field is 
 $g^2\phi^2|X|^2$.  Quantum fluctuations of the $X$ field with
momentum $\vec{k}$ during preheating {\em approximately} obey the
Mathieu equation, $ X_k'' + [A(k) - 2q\cos2z]X_k =0$, where $q = g^2
\phi^2 / 4 M_\phi^2$, $A(k) = (k^2 + M_X^2) / M_\phi^2 + 2q$ (primes
denotes differentiation with respect to $z=M_\phi t$).  Particle
production occurs above the line $A = 2 q$ in an instability strip of
width scaling as $q^{1/2}$ for large $q$.  The condition for broad
resonance, $A-2q \simlt q^{1/2}$ \cite{explosive,klr}, becomes $(k^2 +
M^2_X)/M_\phi^2 \simlt g \bar\phi / M_\phi$, which yields $E_X^2 =
{k^2 + M^2_X } \simlt g \bar\phi M_\phi $ for the typical energy of
particles produced in preheating. Here $\bar\phi$ is the amplitude of
the oscillating inflaton field \cite{explosive}.  The resulting
estimate for the typical energy of particles at the end of the broad
resonance regime for $M_\phi \sim 10^{-6} M_{\rm Pl}$ is $E_X \sim
10^{-1} g^{1/2}\sqrt { M_\phi M_{\rm Pl}} \sim g^{1/2} 10^{15}$ GeV.
Supermassive $X$ bosons can be produced by the broad parametric
resonance for $E_X > M_X$, which leads to the estimate that $X$
production will be possible if $M_X < g^{1/2} 10^{15}$ GeV. For $g^2
\sim 1$ one would have copious production of $X$ particles as heavy as
$10^{15}$GeV, {\it i.e.}, 100 times greater than the inflaton mass,
which may be many orders of magnitude greater than the reheat
temperature. In fact, in an expanding Universe $M_\phi$ and
$\bar{\phi}$ are time-dependent quantities and one should not only
have very large field at the very beginning of the process of
preheating, but also have $E_X^2\simlt g \bar\phi M_\phi $ until the
end of preheating \cite{long}. These considerations lead to an
estimate of the upper bound on $M_X$ slightly smaller than $g^{1/2}
10^{15}$ GeV \cite{long}.  Scatterings of $X$ fluctuations off the
zero mode of the inflaton field considerably limits the maximum
magnitude of $X$ fluctuations to be $\langle X^2\rangle_{\rm max}
\approx M_\phi^2/g^2$ \cite{KT3}.  For example, $\langle
X^2\rangle_{\rm max} \simlt 10^{-10} M_{\rm Pl}^2$ if $M_X =
10\:M_\phi$. This restricts the corresponding number density of
created $X$-particles.

For a reheating temperature of the order of 100 GeV, the present
abundance of SDM with mass $M_X\sim 10^{14}$ GeV is 
$\Omega_{X}\sim 1$ if $\epsilon \sim 10^{-10}$. This small fraction
corresponds to $\langle X^2\rangle \sim 10^{-12} M_{\rm Pl}^2$ at the
end of the preheating stage, a value naturally achieved for SDM masses
in the GUT range \cite{KT3}. The creation of SDM through preheating
and, therefore, the prediction of the present value of $\Omega_X$, is
very model dependent. 
The evolution of the background inflaton field responsible for the $X$
production will be determined by its coupling to other fields since
only a negligible fraction of its energy can go into SDM.  We feel very
encouraged, however, that it is possible to produce supermassive
particles during preheating that are as massive as $10^{12}T_{RH}$.

Another possibility which has been recently investigated is the
production of very massive particles by gravitational mechanisms
\cite{ckr,igor}. In particular, the desired abundance of SDM may be
generated during the transition from the inflationary phase to a
matter/radiation dominated phase as the result of the expansion of the
background spacetime acting on vacuum fluctuations of the dark
matter field \cite{ckr}. A crucial aspect of inflationary
scenarios is the generation of density perturbations. A related
effect, which does not seem to have attracted much attention, is the
possibility of producing matter fields due to the rapid change in the
evolution of the scale factor around the end of inflation. Contrary to
the first effect, the second one contributes to the homogeneous
background energy density that drives the cosmic expansion, and is
essentially the familiar ``particle production'' effect of
relativistic field theory in external fields.

Very massive particles may be created in a nonthermal state in
sufficient abundance for critical density today by
classical gravitational effect on the vacuum state at the end of
inflation. Mechanically, the particle creation scenario is similar to
the inflationary generation of gravitational perturbations that seed
the formation of large scale structures.  However, the quantum
generation of energy density fluctuations from inflation is associated
with the inflaton field which dominated the mass density of the
universe, and not a generic, sub-dominant scalar field.

If $0.04 \simlt M_{X}/H_e \simlt 2$ \cite{ckr}, where $H_e$ is the
Hubble constant at the end of inflation, DM produced gravitationally
can have a density today of the order of the critical density. This
result is quite robust with respect to the fine details of the
transition between the inflationary phase and the matter-dominated
phase.  The only requirement is that
$(H_e/10^{-6}M_{Pl})^2(T_{RH}/10^9{\rm GeV})\simgt10^{-2}$.  The
observation of the cosmic background radiation anisotropy does not fix
uniquely $H_e$, but using $T_{RH}\simlt\sqrt{M_{Pl}H_e}$, we find that
the mechanism is effective only when $H_e \simgt 10^9$GeV (or,
$M_X\simgt 10^8$GeV).

The distinguishing feature of this mechanism \cite{ckr} is
the capability of generating particles with mass of the order of the
inflaton mass even when the SDM only interacts extremely weakly (or
not at all!) with other particles, including the inflaton.   This
feature makes the gravitational production mechanism quite model
independent and, therefore, more appealing to us than the one
occurring at preheating.

Supermassive particles may also be produced in theories where
inflation is completed by a first-order phase transition \cite{ls}. In
these scenarios, the universe decays from its false vacuum state by
bubble nucleation \cite{guth}.  When bubbles form, the energy of the
false vacuum is entirely transformed into potential energy in the
bubble walls, but as the bubbles expand, more and more of their energy
becomes kinetic and the walls become highly relativistic. Eventually
the bubble walls collide.

During collisions, the walls oscillate through each other \cite{moss}
and the kinetic energy is dispersed into low-energy scalar waves
\cite{moss,wat}.  If these soft scalar quanta carry quantum numbers
associated with some spontaneously broken symmetry, they may even lead
to the phenomenon of nonthermal symmetry restoration \cite{col}.  We
are, however, more interested in the fate of the potential energy of
the walls, $M_P = 4\pi\eta R^2$, where $\eta$ is the energy per unit
area of the bubble with radius $R$.  The bubble walls can be imagined
as a coherent state of inflaton particles, so that the typical energy
$E$ of the products of their decays is simply the inverse thickness of
the wall, $E\sim \Delta^{-1}$. If the bubble walls are highly
relativistic when they collide, there is the possibility of quantum
production of nonthermal particles with mass well above the mass
 of the inflaton field, up to energy $\Delta^{-1}=\gamma
M_\phi$, $\gamma$ being the relativistic Lorentz factor.

Suppose now that the SDM is some fermionic degree of freedom $X$ and
that it couples to the inflaton field by the Yukawa coupling $g
\phi\overline{X}{X}$. One can treat $\phi$ (the bubbles or walls) as a
classical, external field and the SDM as a quantum field in the
presence of this source. We are thus ignoring the backreaction of
particle production on the evolution of the walls, but this is
certainly a good approximation in our case. The number of SDM
particles created in the collisions from the wall's potential energy
is $N_X\sim f_X M_P/M_X$, where $f_X$ parametrizes the fraction of the
primary decay products that are supermassive DM particles.  The
fraction $f_X$ will depend in general on the masses and the couplings
of a particular theory in question.  For the Yukawa coupling $g$, it
is $ f_X \simeq g^2 {\rm ln}\left(\gamma M_\phi/2 M_{X}\right)$
\cite{wat,mas}.  Supermassive particles in bubble collisions are
produced out of equilibrium and they never attain chemical
equilibrium. Assuming $T_{RH}\simeq 100$ GeV, the present abundance of
SDM is $\Omega_{X}\sim 1$ if $g\sim 10^{-5}\alpha^{1/2}$. Here
$\alpha^{-1}\ll 1$ denotes the fraction of the bubble energy at
nucleation which has remained in the form of potential energy at the
time of collision.  This simple analysis indicates that the correct
magnitude for the abundance of $X$ particles may be naturally obtained
in the process of reheating in theories where inflation is terminated
by bubble nucleation.

In conclusion, we have suggested  that  DM 
 may be elementary  supermassive dark matter. Its mass
may greatly exceed the electroweak scale, perhaps as the GUT
scale. This is possible because the SDM was created in a
nonthermal state and has never reached chemical equilibrium, 
thus avoiding the unitarity upper bound of about
$10^2$ TeV.  We have reviewed a number of ways SDM may be created. 
If reheating after inflation is
preceded by a preheating stage, it is certainly possible to produce by
resonance effects copious amounts of dark matter particles much
heavier than the inflaton mass. We have also argued that the same may
occur if inflation is completed by a first-order phase transition.
These two scenarios are based on several assumptions about the
structure of the theory, coupling constants, and the reheating
temperature, but it is comfortable that the desirable abundance of
nonthermal massive relics may be generated. Nonthermal SDM may be also
created gravitationally at the end of inflation, with a significant
mass range for which the SDM particles will have critical density
today regardless of the fine details of the inflation-matter/radiation
transition. This production mechanism involves the
dynamics between the classical gravitational field and a quantum
field; it needs no fine tuning of field couplings or any coupling to
the inflaton field.  We are excited that the recent developments in
understanding how matter is created in the early universe suggests
that DM might be supermassive and resurrect the possibility that it
might be charged or even strongly interacting. The implications of
these fascinating options will be discussed elsewhere \cite{inprep}.

DJHC and EWK were supported by the DOE and NASA under Grant NAG5-7092.

\end{document}